# Information Flow of quantum states interacting with closed timelike curves

T.C.Ralph and C.R.Myers

Centre for Quantum Computer Technology, Department of
Physics, University of Queensland, Brisbane 4072, QLD, Australia

(Dated: November 30, 2018)

Recently, the quantum information processing power of closed timelike curves have been discussed. Because the most widely accepted model for quantum closed timelike curve interactions contains ambiguities, different authors have been able to reach radically different conclusions as to the power of such interactions. By tracing the information flow through such systems we are able to derive equivalent circuits with unique solutions, thus allowing an objective decision between the alternatives to be made. We conclude that closed timelike curves, if they exist and are well described by these simple models, would be a powerful resource for quantum information processing.

The fact that general relativity appears to permit the existence of closed timelike curves – time machines that allow a system to interact with its own past [1, 2] – has motivated discussion of how the rules of quantum mechanics might be altered by their existence [3], [4], [5], [6], [7]. Perhaps the most radical conclusion from these studies is that of Brun et al [7], that closed timelike curves (CTCs) can be used to discriminate nonorthogonal states – thus effectively negating the uncertainty principle. However, soon after this suggestion was made, Bennett et al [8] argued that this claim is incorrect and that effectively CTCs erase information that interacts with them.

The results of both sets of authors are based on the model of CTCs introduced by Deutsch [3]. That these authors can arrive at opposite conclusions based on the same model illustrates that there are a number of ambiguities associated with Deutsch's approach, and so we may be tempted to treat both results with some scepticism. In this paper we clarify the situation by rederiving the dynamics of the model by careful tracking the flow of information through the system. Given the assumption that CTCs exist and can be interacted with in the way described by this simple model, our results confirm those of Brun et al that non-orthogonal states can be discriminated.

The model of CTCs introduced by Deutsch [3] runs as follows. Consider the situation depicted in Fig.1(a). A qubit (1) enters in state $\rho_{in} = |\phi\rangle\langle\phi|$ and unitarily interacts with another qubit (2) in state $\rho$. The qubit 1 then enters a wormhole [1] that takes it back in time where it becomes qubit 2, thus creating a CTC. It again passes through the unitary interaction and then propagates out of the region of the wormhole. This depiction of events is slightly rearranged from the usual one. This is done in order to more naturally discuss the flow of information, but is completely equivalent to the usual depiction up to a SWAP operation in the unitary. The Deutsch solution is to firstly determine the state of qubit 2, given by the density operator $\rho$, via the consistency equation

$$\rho = Tr_2[U(\rho_{in} \otimes \rho)U^\dagger] \qquad (1)$$

where the trace is over the Hilbert space of qubit 2. Given a solution for $\rho$, the output state of qubit 2, $\rho_{out}$, is given by

$$\rho_{out} = Tr_1[U(\rho_{in} \otimes \rho)U^\dagger] \qquad (2)$$

where the trace is now over the Hilbert space of qubit 1. In general the solution for $\rho_{out}$ will be a non-unitary and a nonlinear function of the input state.

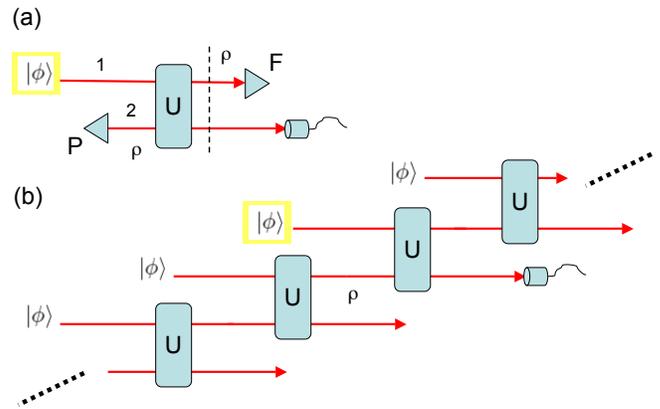

FIG. 1: Model of a qubit interacting with a closed timelike curve formed by a wormhole. (a) As seen by an observer far away from the wormhole. The qubit suffers an elastic collision with a second qubit described by the unitary $U$. The qubit can be scattered into the future mouth of the wormhole (F). It emerges from the past mouth of the wormhole (P), in the past, and becomes the second qubit in the elastic collision, thus forming a closed timelike curve. It can then be scattered into an output mode that is subsequently detected. (b) As "seen" by the qubit. Either propagating forward along the path of the qubit from the initial state or back along the path of the qubit from the detector leads to infinite strings of identical interactions with copies of the qubit. Multiple interactions with the unitary represent cases where the qubit is repeatedly scattered through the wormhole. This equivalent circuit can be solved in the limit of many interactions to give a unique solution to the CTC interaction.

There are two ambiguities in this model:

(1) Although it is guaranteed that for every $U$ and $\rho_{in}$ a solution exists for $\rho$, that solution may not be unique.

If multiple solutions exist Deutsch suggests choosing the one for which $\rho$ is of maximum entropy. However, the physical justification for such a choice is unclear and subsequent authors have considered the question unresolved [5].

(2) Deutsch's original derivation considered pure input states. Because of the nonlinearity inherent in the CTC system the usual equivalence between different ways of describing mixed quantum states does not necessarily apply and can lead to physically distinct predictions.

In the examples considered by Brun et al there are always unique solutions for $\rho$, hence (1) is not a problem. Instead, the repudiation of the result of Brun et al by Bennett et al is based on ambiguity (2). In particular Bennett el al allow $\rho_{in}$ to represent a classically mixed ensemble and apply Deutsch's rules to calculate its evolution. This leads to the claim that if classical correlations exist between qubit 1 and some other qubit, 3, then these correlations may be completely erased by the CTC. This can occur because of the traces taken in Eq.1 and 2. From this it follows that in the Brun et al circuit there are no correlations whatsoever between what is sent into the CTC and what emerges, meaning that no states can be differentiated using the CTC – let alone non-orthogonal states.

In contrast, Brun et al's results follow from calculating the evolution of pure input sub-ensembles and then introducing the classical mixture by taking a mixture over the resulting output states. Doing this preserves the classical correlations and leads to the result that nonorthogonal states can be differentiated. To unambiguously answer the question of how to treat classical correlations we need to re-derive the Deutsch result in a way that explicitly follows the information flow of the system. We will show that doing this in fact removes both ambiguities of the model.

Consider Fig.1(b). Here we have expanded out the propagation of Fig.1(a) into an equivalent circuit. This equivalent circuit can be obtained heuristically by considering what the qubit "sees" when propagating forward through the wormhole or alternatively, by tracing its trajectory back through the wormhole into the past. In both cases the qubit experiences multiple interactions via the unitary $U$ with copies (i.e. perfect clones) of itself. The effective circuit stretches indefinitely into the future and into the past representing the time lines experienced by the qubit, however in the physical circuit (Fig.1(a)) all these paths overlap in the same time interval. We have assumed that the action of the wormhole in creating the CTC is purely geometric, as described by general relativity. The equivalent circuit uniquely describes that geometry as viewed from the perspective of the qubit. The behaviour of the qubits on the equivalent circuit are then assumed to follow the normal rules of quantum physics.

The interpretation of the effective circuit is clarified by considering the limit in which there is no interaction with the wormhole. This occurs when $U = SWAP$. In order to recover standard quantum mechanics in this limit we must assume that all the effective modes except the one indicated to be striking the detector, are lost (i.e. not measured). The other modes can be interpreted as additional degrees of freedom that are not normally observed but can be indirectly probed via the interaction $U$ in the presence of the CTC created by the wormhole. Calculating the Heisenberg evolution of the effective circuit of Fig.1(b) leads to the same expectation values as derived in [6], where the additional degrees of freedom are associated with the space-time geometry.

Here we solve for the Schroedinger evolution. We can solve the effective circuit by starting far in the "past" with the initial state for both arms $\rho_{in}$. After one iteration we obtain $\rho' = Tr_L[U(\rho_{in} \otimes \rho_{in})U^\dagger]$ where the lost (lower) arm has been traced out. After two iterations we have $\rho'' = Tr_L[U(\rho_{in} \otimes \rho')U^\dagger]$. If this map converges to a fixed point then it will be true after many iterations that $\rho = Tr_L[U(\rho_{in} \otimes \rho)U^\dagger]$. This expression coincides with that given by Deutsch for calculating the corresponding $\rho$ in Fig.1(a) [3] (Eq.1). Deutsch showed that such a fixed point always exists, so the assumed convergence is guaranteed. Further, given that the part of the effective circuit that propagates into the "future" is also lost, then the output state is given by $\rho_{out} = Tr_1[U(\rho_{in} \otimes \rho)U^\dagger]$ where now the trace is over the upper qubit. Again this corresponds to the expression given by Deutsch for calculating the output state in Fig.1(a) [3] (Eq.2). Thus the circuit of Fig.1(b) is mathematically equivalent to that of Fig.1(a).

An exception to this rule are cases where there exist multiple fixed points. The iterative solution of the equivalent circuit nonetheless converges to a unique solution, thus removing the first ambiguity of the Deutsch approach. But is this unique solution the one suggested by Deutsch? To answer this we must insert a little more physics into the problem. A general characteristic of situations in which multiple solutions exist for the Deutsch approach is that quantum information becomes trapped for arbitrarily long times, endlessly cycling through the CTC (see appendix). Under such circumstances it is unphysical to assume that the interaction, $U$, is perfectly unitary. If an arbitrarily small, but finite amount of decoherence is added to $U$, circuits with unique solutions remain essentially unchanged. However, circuits with multiple fixed points converge to the unique fixed point of the maximum entropy $\rho$. Thus we find that the equivalent circuit approach resolves the issue of the multiple solutions ambiguity in agreement with Deutsch's original suggestion. The insight is that any physical quantum interaction will involve some non-zero (even if very small) coupling to the environment. When this is included, the unique solution is always the one corresponding to maximum entropy as is proved in the appendix.

To answer the second ambiguity in the Deutsch approach we need to represent the process that produces the input state in the most general way. The input state for a single shot experiment can always be written as a pure state provided all the modes involved in producing it are explicitly included. This is represented in Fig.2 where the qubit mode is coupled to many environmental modes via a unitary interaction. The macroscopic arrangement of the apparatus which produces the interaction is labelled by the classical parameters. The result of the interaction is to produce a pure state $|\phi\rangle$ of the combined qubit - environment system. Suppose that we now take this more general state as the input to our system in Fig.1(a). In constructing the equivalent circuit we note again that the qubit "sees" many copies of this identical physical arrangement leading to many copies of $\rho_{in} = |\phi\rangle\langle\phi|$. That is, the composite input state for the equivalent circuit is $\rho_{inE} = |\phi\rangle^{\otimes n}\langle\phi|^{\otimes n}$, where formally $n \to \infty$. These multi-mode pure state inputs were also considered by Deutsch and our circuit expansion remains equivalent. There is no ambiguity here as the tracing out of the environmental modes needed to obtain the final answer can be performed before or after the CTC interaction, delivering the same answer by either approach.

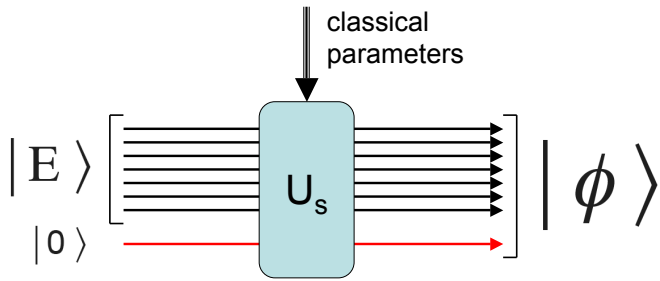

FIG. 2: Generic description of state production where, for a single shot, the global output state, $|\phi\rangle$, is always pure provided sufficient ancillary environmental modes are included. The initial state of the qubit is $|0\rangle$ and that of the environmental modes is $|E\rangle$. The state of the qubit is obtained by tracing over the environmental modes. The process that produces the output is modeled as a unitary $U_s$. The macroscopic arrangement of the apparatus producing the process is modeled by the classical parameters determining the unitary.

Now suppose instead of a single shot, we consider how to represent the input state when an ensemble is considered, and particularly when the classical parameters determining the macroscopic arrangement of the apparatus vary randomly shot to shot through the ensemble. Whilst this question is ambiguous to answer in the Deutsch approach, it is a trivial generalization of the equivalent circuit – we simply form a mixed ensemble in the usual way from the single shot input states. Suppose there are $K$ different classical setting in the ensemble each producing the single shot state $|\phi_k\rangle$ with weighting $P_k$. Then the equivalent circuit's composite input state for the mixed ensemble is

$$\rho_{inE} = \Sigma_{k=1}^{K} P_k |\phi_k\rangle^{\otimes n}\langle\phi_k|^{\otimes n} \qquad (3)$$

This now removes the second ambiguity in the Deutsch approach as it uniquely defines how to treat a classically mixed input state. In particular it agrees with the answer given by mixing over the outputs of pure sub-ensembles given by the Deutsch approach, not that given by mixing over the input.

We need now consider one final generalization. That is, the case in which the input state is bipartite, and the arm that does not interact with the CTC is kept and correlated with the arm that did. This situation is depicted in Fig.3(a) and the equivalent circuit is shown in Fig.3(b). As for the simpler single mode case we can clarify the interpretation of the equivalent circuit by requiring that when $U = SWAP$ we recover standard quantum mechanics. This identifies the two modes shown striking detectors as those that are observed. All other modes are lost and should be traced over.

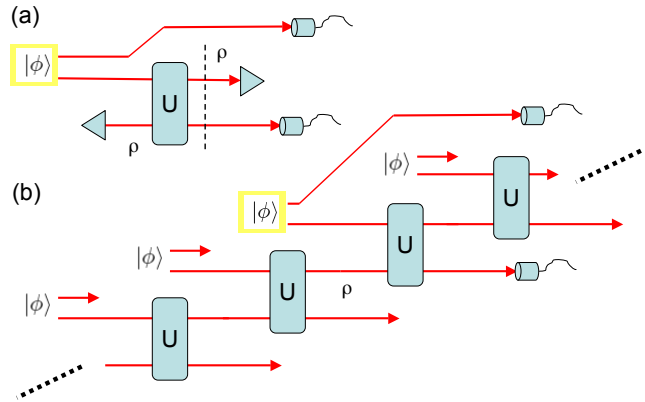

FIG. 3: Model of a bipartite qubit state where one of the subsystems is interacting with a closed timelike curve formed by a wormhole. After the interactions both arms are measured. As for Fig.1, (a) shows the view from outside whilst (b) represents the qubits view. Considering the case where the qubits don't interact with the wormhole at all (i.e. when $U = SWAP$) uniquely identifies the modes shown striking detectors as the only ones observed in the equivalent circuit (b).

Consider now the case in which $|\phi\rangle$ is a maximally entangled state and $U = I$, i.e. one arm of the entanglement travels through the wormhole but there is no interaction. From the equivalent circuit we see that the observed modes originate from different copies of the entangled state and thus are completely uncorrelated. This decorrelation of entanglement was discussed by Deutsch and is what ensures the non-linearity of the CTC can not be used for faster than light communication [3]. Now consider the case of classical correlations – in particular suppose half the time the state $|0\rangle|0\rangle$ is produced and the other half $|1\rangle|1\rangle$ is produced. Using the equivalent circuit with the initial state

$$\rho_{inE} = 1/2(|0\rangle|0\rangle)^{\otimes n}(\langle 0|\langle 0|)^{\otimes n} + 1/2(|1\rangle|1\rangle)^{\otimes n}(\langle 1|\langle 1|)^{\otimes n} \qquad (4)$$

we simply need to pick out the observed modes and trace over the rest. This is trivial and leads to

$$\rho_{out} = 1/2|0\rangle|0\rangle\langle 0|\langle 0| + 1/2|1\rangle|1\rangle\langle 1|\langle 1| \quad (5)$$

showing that, unlike quantum correlations, classical correlations are not destroyed by the wormhole.

The physical insight here is that classical correlations are determined by local hidden variables, i.e. the classical parameters that determine the configuration of the experimental apparatus. The CTC can create copies of these parameters because they are available locally. In contrast an entangled state can not be described by such local hidden variables. As the CTC only "samples" one arm of the entangled state it cannot reproduce the nonlocal correlations of the other arm. Bennett et al [8] call the ambiguity with respect to mixed states in Deutsch's formalism the "linearity trap" and suggest it should be resolved by insisting that "the evolution of a nonlinearly evolving system may depend on parts of the universe with which it does not interact". Our analysis of the information flow of systems that include CTCs comes to the opposite and more pragmatic conclusion that the evolution of a nonlinearly evolving system may only depend on parts of the universe with which it directly interacts. In particular, non-local correlations, as produced by entanglement, do not correspond to direct interactions. We note that this is more in keeping with the local character of general relativity.

We can now use the effective circuit to rigorously compute Brun et al's examples. The simplest interaction is one in which $U$ is a controlled-Hadamard gate (CH), with the control on the lower qubit, and we attempt to differentiate the states $|0\rangle$ and $|-\rangle$. Following Bennett et al's prescription [8] we have Victor prepare the mixed ensemble $1/2|0\rangle_v|0\rangle\langle 0|\langle 0|_v + 1/2|1\rangle_v|-\rangle\langle -|\langle 1|_v$. He keeps one qubit (labelled $v$) so that he can remember which state was prepared and hands the other qubit to Alice, who uses the CTC, with $U = CH$, to analyse it. As previously described the correct input state for the equivalent circuit is then

$$\begin{aligned}\rho_{inE} &= 1/2(|0\rangle_v|0\rangle)^{\otimes n}(\langle 0|\langle 0|_v)^{\otimes n} \\ &+ 1/2(|1\rangle_v|-\rangle)^{\otimes n}(\langle -|\langle 1|_v)^{\otimes n}\end{aligned} \quad (6)$$

Calculating the relevant outputs from the equivalent circuit leads to the following shared state between Victor and Alice after the processing by the CTC

$$\begin{aligned}\rho_{out} &= 1/2|0\rangle_v|0\rangle\langle 0|\langle 0|_v + 1/2 \\ &\quad |1\rangle_v \lim_{n\to\infty}((1/2)^n|0\rangle\langle 0| + (1-(1/2)^n)|1\rangle\langle 1|) + \\ &\quad (1/2)^{n+1/2}(|1\rangle\langle 0| + |0\rangle\langle 1|))\langle 1|_v \\ &= 1/2|0\rangle_v|0\rangle\langle 0|\langle 0|_v + 1/2|1\rangle_v|1\rangle\langle 1|\langle 1|_v \quad (7)\end{aligned}$$

By measuring her qubit in the computational basis Alice can now predict perfectly whether Victor handed her a zero or anti-diagonnal qubit thus deterministically differentiating non-orthogonnal states as predicted by Brun et al. Similar, though more complicated, calculations also confirm Brun et als more sophisticated circuits.

A criticism that could be leveled at these calculations (and those in References [7] and [8]) is that they use non-relativistic quantum mechanics in spite of the fact that wormholes and CTCs are highly relativistic objects. Initial steps towards placing such calculations into a relativistic frame work have been taken in References [6] and [9]. These approaches are consistent with the equivalent circuit formalism introduced here. In particular, because the equivalent circuits are physical circuits, the inclusion of qubit dynamics or more general field states and interactions is straightforward.

The equivalent circuits we have introduced could be approximately constructed in the laboratory, thus allowing experimental simulations of interesting CTC interactions.

In this paper we have considered the information flow of quantum evolutions occurring in the presence of CTCs by deriving and solving equivalent circuits. We have proven that these circuits lead to identical solutions to those of the Deutsch approach in all unambiguous situations, but lead to unique solutions for cases where the Deutsch approach is ambiguous. Our work supports the conclusions of Brun et al that an observer can use interactions with a CTC to allow them to discriminate unknown, non-orthogonal quantum states – in contradiction of the uncertainty principle.

We acknowledge useful discussions with Nick Menicucci, Gerard Milburn and Jacques Pienaar. This work was supported by the Australian Research Council and the Defence Science and Technology Organization.

### A. Appendix

In this appendix we prove that the unique solution obtained from the equivalent circuit approach agrees with the maximum entropy solution proposed by Deutsch [3], provided we allow for the fact that any physical interaction will necessarily involve some non-zero level of decoherence. We proceed in three steps: (i) we show that non-unique solutions in the Deutsch model correspond to equivalent circuits that are sensitive to initial conditions; (ii) we show that this sensitivity is removed when any non-zero amount of decoherence is added to the interaction; and (iii) the resulting unique solution has maximum entropy with respect to all other possible solutions, as conjectured by Deutsch [3].

(i) In the main text we solved for the equivalent circuit by taking an initial state for which both arms were in the state $\rho_{in}$. Here we consider a more general case and take the state of the lower arm to be some arbitrary state $\rho_o$. For a particular choice of $\rho_{in}$ and $\rho_o$ the equivalent circuit will reach a unique fixed point. However, we can consider two different cases: (a) the fixed point $\rho(\rho_{in})$ does not depend on the choice of $\rho_o$. In this case the fixed point is uniquely determined by $\rho_{in}$ and there will be only one solution to the Deutsch consistency condition; or (b) the fixed point $\rho(\rho_{in}, \rho_o)$ does depend on the choice of $\rho_o$. Now there are multiple possible solutions depending on the particular choice of $\rho_o$. This case will lead to multiple solutions for the Deutsch consistency condition (for example see [5]).

(ii) For case (b) we can view the equivalent circuit as a quantum channel that takes an input $\rho_o$ to an output $\rho(\rho_{in}, \rho_o)$ via a long chain of identical processes $\epsilon(\rho_{in})$ (see Fig 1(a)). Suppose now a small but finite amount of isotropic decoherence is added as shown in Fig.1(b). For an arbitrarily long channel the output $\rho_d(\rho_{in})$, will no longer depend on $\rho_o$. This follows from the fact that if $\epsilon$ was the identity, then a sufficiently long channel would be completely depolarising and would send all inputs to the identity, i.e. $\rho_o \to I$. The only way information about $\rho_o$ could be preserved would be if the channel was error correcting. However the first requirement of error correction is that the input is encoded into code words with specific properties [10]. As $\rho_o$ is not encoded the channel cannot be error correcting and so all information about $\rho_o$ will be erased from the output regardless of the nature of $\epsilon$. Therefore, the inclusion of a small amount of decoherence inevitably leads to a solution that cannot depend on $\rho_o$ and hence is unique, even if multiple solutions exist for the Deutsch consistency condition.

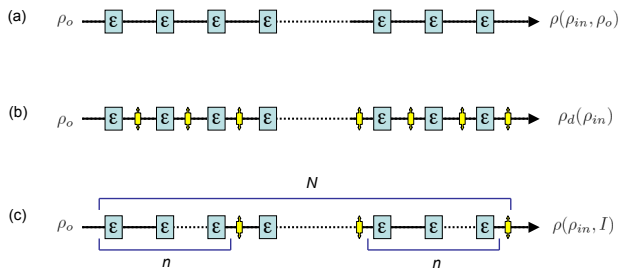

FIG. 4: Depiction of the equivalent circuit as a quantum channel consisting of: (a) many identical processes $\epsilon$; (b) with the addition of decohering elements; and (c) where we consider blocks of $n$ processes that are sufficient in number to lead to convergence to the fixed point but for which decoherence can be neglected. Never-the-less for $N$ processes ($N >> n$) decoherence is significant.

(iii) It now remains to prove that the output $\rho_d(\rho_{in})$ is equal to $Max[\rho(\rho_{in}, \rho_o)]$, where we are maximising the entropy as a function of $\rho_o$. We proceed by introducing two timescales to the problem (see Fig 1(c)). We assume that after $n$ iterations of $\epsilon$ the channel has converged to its fixed point, but that on this scale decoherence is negligible. However, after a much larger number of iterations $N >> n$, decoherence is significant. This justifies adding a decohering element only after each block of $n$ interactions. After the first block of $n$ iterations we have the output $\rho(\rho_{in}, \rho_o)$. Adding a small amount of depolarisation leads to the transformation [11]

$$\rho(\rho_{in}, \rho_o) \to (1-p)\rho(\rho_{in}, \rho_o) + p\, I, \quad (8)$$

where $p << 1$ is the rate of depolarisation. Thus after the second block of $n$ interactions the output will be $(1-p)\rho(\rho_{in}, \rho_o) + p\,\rho(\rho_{in}, I)$. Iterating this process $N$ times leads to the output

$$(1-p)^N \rho(\rho_{in}, \rho_o) + (1-(1-p)^N)\rho(\rho_{in}, I). \quad (9)$$

For sufficiently large $N$ this converges to $\rho(\rho_{in}, I)$. Finally, we need to show that $\rho(\rho_{in}, I) = Max[\rho(\rho_{in}, \rho_o)]$. To do this first write $\rho(\rho_{in}, \rho_o)$ in terms of a Kraus decomposition of the process [11]

$$\rho(\rho_{in}, \rho_o) = \Sigma_k E_k \rho_o E_k^\dagger \quad (10)$$

where the Kraus operators $E_k$ satisfy

$$\Sigma_k E_k^\dagger E_k = I \quad (11)$$

Because $\rho(\rho_{in}, \rho_o)$ is a fixed point, it further follows that

$$\Sigma_k E_k \rho_o E_k^\dagger = \Sigma_k E_k (\Sigma_j E_j \rho_o E_j^\dagger) E_k^\dagger \quad (12)$$

Combining Eqs 11 and 12, we can write

$$\Sigma_k E_k (\Sigma_j E_j^\dagger E_j) \rho_o E_k^\dagger = \Sigma_k E_k (\Sigma_j E_j \rho_o E_j^\dagger) E_k^\dagger \quad (13)$$

Eq 13 can only be satisfied if $[E_j \rho_o, E_j^\dagger] = 0$. Therefore Eq 10 can be rewritten as

$$\rho(\rho_{in}, \rho_o) = \Sigma_k E_k^\dagger E_k \rho_o \quad (14)$$

and hence we conclude that $\rho(\rho_{in}, I) = I$. As $I$ is globally the maximum entropy state then we have $\rho(\rho_{in}, I) = Max[\rho(\rho_{in}, \rho_o)]$ as required.

In summary, we have shown that the inclusion of an arbitrarily small, but non-zero amount of decoherence in the equivalent circuit leads it to converge to a unique solution that corresponds to the solution found from Deutsch's consistency condition with the added maximum entropy requirement. The same solution is obtained from the equivalent circuit by considering idealized unitaries but starting the iteration from the identity.